\documentclass[twocolumn,aps,pra,showpacs]{revtex4}

\usepackage{graphicx}

\begin{document}
\title{Greenberger-Horne-Zeilinger-like proof of Bell's theorem
involving observers who do not share a reference frame}
\author{Ad\'{a}n Cabello}
\email{adan@us.es} \affiliation{Departamento de F\'{\i}sica
Aplicada II, Universidad de Sevilla, 41012 Sevilla, Spain}
\date{\today}


\begin{abstract}
Vaidman described how a team of three players, each of them
isolated in a remote booth, could use a three-qubit
Greenberger-Horne-Zeilinger state to always win a game which would
be impossible to always win without quantum resources. However,
Vaidman's method requires all three players to share a common
reference frame; it does not work if the adversary is allowed to
disorientate one player. Here we show how to always win the game,
even if the players do not share any reference frame. The
introduced method uses a 12-qubit state which is invariant under
any transformation $R_a \otimes R_b \otimes R_c$ (where $R_a = U_a
\otimes U_a \otimes U_a \otimes U_a$, where $U_j$ is a unitary operation
on a single qubit) and requires only single-qubit measurements. A
number of further applications of this 12-qubit state are
described.
\end{abstract}


\maketitle


\section{Introduction}
\label{Sec1}


In 1991, after months of patient ``work'' and based on a study of
$20\,000$ events, a gang of players reached an amazing conclusion:
in eight roulette wheels of the Gran Casino of Madrid, six numbers
($1$ and its two neighbors, $20$ and $33$, and the opposite number
in the roulette wheel, $4$, and its two neighbors, $19$ and $21$)
occurred with an unexpectedly high frequency (assuming that each
of the $37$ numbers of the roulette wheel appears with the same
frequency), while four numbers ($11$, $12$, $28$, and $36$) rarely
occurred. The gang won a large amount of money by betting in these
roulette wheels. The casino never realized where the problem was,
never understood the ``method'' used by the gang but, after many
attempts, found its own method to defeat the gang: the casino
started to regularly exchange the pieces of the roulette wheels
and switch the numbers' positions. This altered the roulette
wheels' original ``defects'' and the gang stopped
winning~\cite{GG03}. The moral is that any winning strategy
usually has an antidote.

In 1999, Vaidman~\cite{Vaidman99} converted
Mermin's~\cite{Mermin90a,Mermin90b} version of the proof of Bell's
theorem without inequalities discovered by Greenberger, Horne, and
Zeilinger (GHZ)~\cite{GHZ89,GHZ90,GHSZ90} into a game involving a
team (a gang) of three players, each of them completely isolated
in a booth, and an opponent (a casino). Under some assumptions,
and using only classical resources, the maximum probability for
the team to win Vaidman's game is 75\% (thus a casino gets profit
by exploiting the remaining 25\%). Thanks to the fact that rules
of the game do not forbid the players to share qubits prepared in
some entangled state, there is a method which allows them to
always win the game. However, there is a simple manipulation that
nullifies the quantum advantage. A hidden assumption of the method
is that all three players share a common reference frame. If the
casino disorientates one of the players so that all three of them
do not share a reference frame, then the advantage of the method
is lost. The term ``unspeakable information'' was coined by Peres
and Scudo~\cite{PS02} to designate information that cannot be
represented by a sequence of discrete symbols, such as a direction
in space or a reference frame. In this paper we show that there is
a method to always win Vaidman's game without it being necessary
that the players share unspeakable information.

In Sec.~\ref{Sec2} we review the rules of Vaidman's game and the
original quantum method for always winning. In Sec.~\ref{Sec3} we
propose a quantum method for always winning, even if the players
do not share any reference frame. This method requires more
qubits, and thus one might think that it must require collective
measurements on several qubits, instead of single-qubit
measurements, as in the original method; in Sec.~\ref{Sec4} we
shall see that this is not the case. In Sec.~\ref{Sec5} we show
other applications of the method.


\section{Vaidman's game}
\label{Sec2}


\subsection{Rules}


Vaidman proposed the following game~\cite{Vaidman99}. Consider a
team of three players, who are allowed to agree on a common
strategy and make any preparation before they are taken to three
remote and isolated booths. Then, each player is asked one of the
two possible questions: ``What is $Z$?'' or ``What is $X$?'' Each
player must give an answer which is limited to one of only two
possibilities:~``$0$'' or~``$1$.'' One of the rules of the game is
that either all three players are asked the~$Z$ question or only
one player is asked the~$Z$ question and the other two are asked
the~$X$ question. The team wins if the number of~$0$ answers is
odd (one or three) in the case of three~$Z$~questions, and is even
(zero or two) in the case of one~$Z$ and two~$X$~questions.

Assuming that the four possible combinations of questions (i.e.,
$Z_1,Z_2,Z_3$; $Z_1,X_2,X_3$; $X_1,Z_2,X_3$; and~$X_1,X_2,Z_3$)
are asked with the same frequency, no classical protocol allows
the players to win the game in more than~75\% of the runs. For
instance, a simple strategy that allows them to win in ~75\% of
the runs is that each player always answers~$1$ to the~$Z$
question and~$0$ to the~$X$ question. However, quantum
mechanics provides a method to always win the game.


\subsection{GHZ-assisted quantum always winning strategy}


The method for always winning is the following. Before entering
the isolated booths, the players prepare a large number of
three-qubit systems in the GHZ
state~\cite{Mermin90a,Mermin90b,GHZ89,GHZ90,GHSZ90,Svetlichny87}
\begin{equation}
|{\rm GHZ}\rangle = {1 \over \sqrt{2}}
(|y_0,y_0,y_0\rangle+|y_1,y_1,y_1\rangle). \label{GHZ03}
\end{equation}
Here $|y_0,y_0,y_0\rangle = |y_0\rangle \otimes |y_0\rangle
\otimes |y_0\rangle$, where $|y_0\rangle = {1 \over \sqrt{2}}
(|z_0\rangle + i|z_1\rangle)$ and $|y_1\rangle = {1 \over
\sqrt{2}} (|z_0\rangle - i|z_1\rangle)$, where $|z_0\rangle =
\left(
\begin{array}{c} 1 \\ 0
\end{array} \right)$ and $|z_1\rangle = \left( \begin{array}{c} 0 \\ 1
\end{array} \right)$.
Then, for each three-qubit system, each of the players takes one
of the qubits with him. In case a player is asked ``What is
$Z$?,'' he performs a measurement on his qubit of the observable
represented by
\begin{equation}
Z=|z_0\rangle \langle z_0|-|z_1\rangle \langle z_1|,
\end{equation}
and gives the answer~$0$, if the outcome corresponds
to~$|z_0\rangle$, or the answer~$1$, if the outcome corresponds
to~$|z_1\rangle$.

In case a player is asked ``What is~$X$?,'' he performs a
measurement of the observable represented by
\begin{equation}
X=|x_0\rangle \langle x_0|-|x_1\rangle \langle x_1|,
\end{equation}
where~$|x_0\rangle = {1 \over \sqrt{2}} (|z_0\rangle +
|z_1\rangle)$ and $|x_1\rangle = {1 \over \sqrt{2}} (|z_0\rangle -
|z_1\rangle)$, and gives the answer~$0$, if the outcome
corresponds to~$|x_0\rangle$, or the answer~$1$, if the outcome
corresponds to~$|x_1\rangle$.

The protocol described above allows the team to always win the
game, because the state defined in Eq.~(\ref{GHZ03}) can also be
expressed in the following four forms:
\begin{eqnarray}
|{\rm GHZ}\rangle & = & {1 \over 2} (|z_0,z_0,z_0\rangle-|z_0,z_1,z_1\rangle \nonumber \\
& & -|z_1,z_0,z_1\rangle-|z_1,z_1,z_0\rangle)
\label{zzz} \\
& = & {1 \over 2} (|z_0,x_0,x_1\rangle+|z_0,x_1,x_0\rangle \nonumber \\
& & -|z_1,x_0,x_0\rangle+|z_1,x_1,x_1\rangle)
\label{zxx} \\
& = & {1 \over 2} (|x_0,z_0,x_1\rangle-|x_0,z_1,x_0\rangle \nonumber \\
& & +|x_1,z_0,x_0\rangle+|x_1,z_1,x_1\rangle)
\label{xzx} \\
& = & {1 \over 2} (-|x_0,x_0,z_1\rangle+|x_0,x_1,z_0\rangle \nonumber \\
& & +|x_1,x_0,z_0\rangle+|x_1,x_1,z_1\rangle).
\label{xxz}
\end{eqnarray}
It can be inferred from Eq.~(\ref{zzz}) that, if all players
measure~$Z$, then either all of them will obtain~$z_0$ or one will
obtain~$z_0$ and the other two will obtain~$z_1$. Analogously, it
can be inferred from Eqs.~(\ref{zxx})--(\ref{xxz}) that, if one
player measures~$Z$ and the other two measure~$X$, then either all
of them will obtain~$1$, or one will obtain~$1$ and the other
two will obtain~$0$.


\section{Quantum always winning strategy without unspeakable information}
\label{Sec3}


The method described above has one drawback that the adversary
could use to keep the players from always winning. If the qubits
are spin states of spin-${1 \over 2}$ particles, then the
observables~$Z$ and $X$ can be identified, respectively, with the
spin components along two orthogonal directions $z$ and $x$. Such
directions are determined by the preparation of the GHZ
state~(\ref{GHZ03}). This method requires all players to share the
directions $z$ and $x$ for the duration of the game. However, if
the opponent finds a way to confuse one of them, then the local
measurements performed by the players will not be adequately
correlated and thus the advantage provided by the GHZ state is
lost.

Fortunately, there is a method which is still valid even if the
players do not share two directions. Now, before entering the
booths, the players prepare a large number of 12-qubit systems in
the state
\begin{equation}
| \Psi\rangle={1 \over \sqrt{2}}
(|\eta_0,\eta_0,\eta_0\rangle+|\eta_1,\eta_1,\eta_1\rangle),
\label{GHZ12}
\end{equation}
where
$|\eta_0\rangle={1 \over \sqrt{2}} (|\phi_0\rangle+i
|\phi_1\rangle)$ and $|\eta_1\rangle={1 \over \sqrt{2}}
(|\phi_0\rangle-i |\phi_1\rangle)$, where
$|\phi_0\rangle$ and $|\phi_1\rangle$
are the four-qubit states
\begin{eqnarray}
|\phi_0\rangle & = & {1 \over 2}
(|z_0,z_1,z_0,z_1\rangle-|z_0,z_1,z_1,z_0\rangle
\nonumber \\ & &
-|z_1,z_0,z_0,z_1\rangle +|z_1,z_0,z_1,z_0\rangle),
\label{P0}
\\
|\phi_1\rangle & = & {1 \over 2 \sqrt{3}}
(2|z_0,z_0,z_1,z_1\rangle-|z_0,z_1,z_0,z_1\rangle
\nonumber \\ & &
-|z_0,z_1,z_1,z_0\rangle -|z_1,z_0,z_0,z_1\rangle
\nonumber \\ & &
-|z_1,z_0,z_1,z_0\rangle+2|z_1,z_1,z_0,z_0\rangle),
\label{P1}
\end{eqnarray}
introduced by Kempe {\em et al.}~\cite{KBLW01} in the context of
decoherence-free fault-tolerant universal quantum
computation~\cite{ZR97a,LCW98}, and recently obtained
experimentally using parametric down-converted
polarization-entangled photons~\cite{BEGKCW03}.

Then, for each 12-qubit system, the first player takes the first
{\em four} qubits with him, the second player takes the next four
qubits, and the third player takes the last four qubits. In case a
player is asked ``What is $Z$?,'' he performs on his four qubits a
measurement of the observable represented by
\begin{equation}
{\cal Z} = |\phi_0\rangle \langle\phi_0|- |\phi_1\rangle
\langle\phi_1|.
\label{calZ}
\end{equation}
The observable ${\cal Z}$ has {\em three} possible outcomes
(corresponding to its three eigenvalues, $-1$, $0$, and $1$).
However, if the qubits have been prepared in the state
$|\Psi\rangle$ given in Eq.~(\ref{GHZ12}), then only two outcomes can
occur (those corresponding to the eigenvalues $-1$ and $1$).
Measuring the observable ${\cal Z}$ on a system prepared in the
state $|\Psi\rangle$ is then equivalent to reliably discriminating
between the states~$|\phi_0\rangle$ and~$|\phi_1\rangle$. The
player gives the answer~$0$, if the outcome corresponds
to~$|\phi_0\rangle$, and the answer~$1$, if the outcome
corresponds to~$|\phi_1\rangle$.

In case a player is asked ``What is~$X$?,'' he performs a
measurement of the observable represented by
\begin{equation}
{\cal X} = |\psi_0\rangle \langle\psi_0|- |\psi_1\rangle
\langle\psi_1|,
\label{calX}
\end{equation}
where
\begin{eqnarray}
|\psi_0\rangle & = & {1 \over \sqrt{2}}
(|\phi_0\rangle+|\phi_1\rangle), \label{S0}
\\
|\psi_1\rangle & = & {1 \over \sqrt{2}}
(|\phi_0\rangle-|\phi_1\rangle). \label{S1}
\end{eqnarray}
Measuring ${\cal X}$ on a system prepared in the state~$|\Psi
\rangle$ is equivalent to reliably discriminating
between~$|\psi_0\rangle$ and~$|\psi_1\rangle$. The player gives
the answer~$0$, if the outcome corresponds to~$|\psi_0\rangle$,
or the answer~$1$, if the outcome corresponds
to~$|\psi_1\rangle$.

The state $|\Psi\rangle$ can be expressed in the following four
forms:
\begin{eqnarray}
| \Psi\rangle
& = & {1 \over 2} (|\phi_0,\phi_0,\phi_0\rangle-|\phi_0,\phi_1,\phi_1\rangle \nonumber \\
& & -|\phi_1,\phi_0,\phi_1\rangle-|\phi_1,\phi_1,\phi_0\rangle)
\label{ZZZ} \\
& = & {1 \over 2} (|\phi_0,\psi_0,\psi_1\rangle+|\phi_0,\psi_1,\psi_0\rangle \nonumber \\
& & -|\phi_1,\psi_0,\psi_0\rangle+|\phi_1,\psi_1,\psi_1\rangle)
\label{ZXX} \\
& = & {1 \over 2} (|\psi_0,\phi_0,\psi_1\rangle-|\psi_0,\phi_1,\psi_0\rangle \nonumber \\
& & +|\psi_1,\phi_0,\psi_0\rangle+|\psi_1,\phi_1,\psi_1\rangle)
\label{XZX} \\
& = & {1 \over 2} (-|\psi_0,\psi_0,\phi_1\rangle+|\psi_0,\psi_1,\phi_0\rangle \nonumber \\
& & +|\psi_1,\psi_0,\phi_0\rangle+|\psi_1,\psi_1,\phi_1\rangle).
\label{XXZ}
\end{eqnarray}
From Eq.~(\ref{ZZZ}), it can be inferred that if the three players
perform measurements to discriminate between~$|\phi_0\rangle$
and~$|\phi_1\rangle$, then they will always obtain an odd number
of states~$|\phi_0\rangle$. From Eqs.~(\ref{ZXX}) to (\ref{XXZ}),
it can be inferred that, if two players perform measurements to
discriminate between~$|\psi_0\rangle$ and~$|\psi_1\rangle$, and
the third performs measurements to discriminate
between~$|\phi_0\rangle$ and~$|\phi_1\rangle$, then they will
always obtain an odd number of states~$|\psi_1\rangle$
and~$|\phi_1\rangle$.

For our purposes, the fundamental property of the state $|\Psi
\rangle$ is that it is invariant under any transformation $R_a
\otimes R_b \otimes R_c$ (where $R_a = U_a \otimes U_a \otimes U_a
\otimes U_a$, where $U_j$ is a unitary operation on a single
qubit). This property derives from the fact that $|\phi_0\rangle$
and $|\phi_1\rangle$ and any linear combination thereof (such as
$|\psi_0\rangle$ and $|\psi_1\rangle$) are invariant under the
tensor product of four equal unitary operators, $U_j \otimes U_j
\otimes U_j \otimes U_j$. This means that the state $|\Psi
\rangle$ is invariant under local rotations, and the local
observables ${\cal Z}$ and ${\cal X}$ are invariant under $U_j
\otimes U_j \otimes U_j \otimes U_j$ and thus under rotations of
the local setups \cite{Cabello03a}. Therefore,
expressions~(\ref{ZZZ})--(\ref{XXZ}) remain unchanged after local
rotations. This implies that even if the adversary disorientates
one or more players, the outcomes of the local measurements still
possess the desired correlations, because the involved local
measurements are rotationally invariant.


\section{Measuring the observables by using single-qubit measurements}
\label{Sec4}


One might think that measuring ${\cal Z}$ (i.e., distinguishing
between $|\phi_0\rangle$ and $|\phi_1\rangle$) and ${\cal X}$
(i.e., distinguishing between $|\psi_0\rangle$ and $|\psi_1
\rangle$) could require collective measurements on each player's
four qubits. However, as in the original method, only single-qubit
measurements are needed.


\subsection{Distinguishing between $|\phi_0\rangle$ and $|\phi_1\rangle$}


The states $|\phi_0\rangle$ and $|\phi_1\rangle$ are reliably
distinguishable using single-qubit measurements because they can
be expressed as
\begin{eqnarray}
|\phi_0\rangle & = & {1 \over 2}
(-|z_0,z_1,x_0,x_1\rangle+|z_0,z_1,x_1,x_0\rangle \nonumber \\ & &
+|z_1,z_0,x_0,x_1\rangle-|z_1,z_0,x_1,x_0\rangle),
\\
|\phi_1\rangle & = & {1 \over 2 \sqrt{3}}
(|z_0,z_0,x_0,x_0\rangle-|z_0,z_0,x_0,x_1\rangle \nonumber \\ & &
-|z_0,z_0,x_1,x_0\rangle+|z_0,z_0,x_1,x_1\rangle \nonumber \\ & &
-|z_0,z_1,x_0,x_0\rangle+|z_0,z_1,x_1,x_1\rangle \nonumber \\ & &
-|z_1,z_0,x_0,x_0\rangle+|z_1,z_0,x_1,x_1\rangle \nonumber \\ & &
+|z_1,z_1,x_0,x_0\rangle+|z_1,z_1,x_0,x_1\rangle \nonumber \\ & &
+|z_1,z_1,x_1,x_0\rangle+|z_1,z_1,x_1,x_1\rangle).
\end{eqnarray}
Therefore, if the local measurements are $Z_1$ (i.e., the
component along the $z$~direction of the first qubit), $Z_2$
(i.e., the component along the $z$~direction of the second qubit),
$X_3$ (i.e., the component along the $x$~direction of the third
qubit), and $X_4$ (i.e., the component along the $x$~direction of
the fourth qubit) then, among the~$16$ possible outcomes,~$4$
occur (with equal probability) only if the qubits were in the
state $|\phi_0\rangle$, and the other~$12$ outcomes occur (with
equal probability) only if the qubits were in the state
$|\phi_1\rangle$. Note that now $z$ and $x$ are not fixed
directions, but any two orthogonal directions instead. This scheme
to distinguish between $|\phi_0\rangle$ and $|\phi_1\rangle$ using
only single-qubit measurements has recently been experimentally
implemented~\cite{BEGKCW03}.


\subsection{Distinguishing between $|\psi_0\rangle$ and $|\psi_1\rangle$}

\begin{figure}
\centerline{\includegraphics[width=8.6cm]{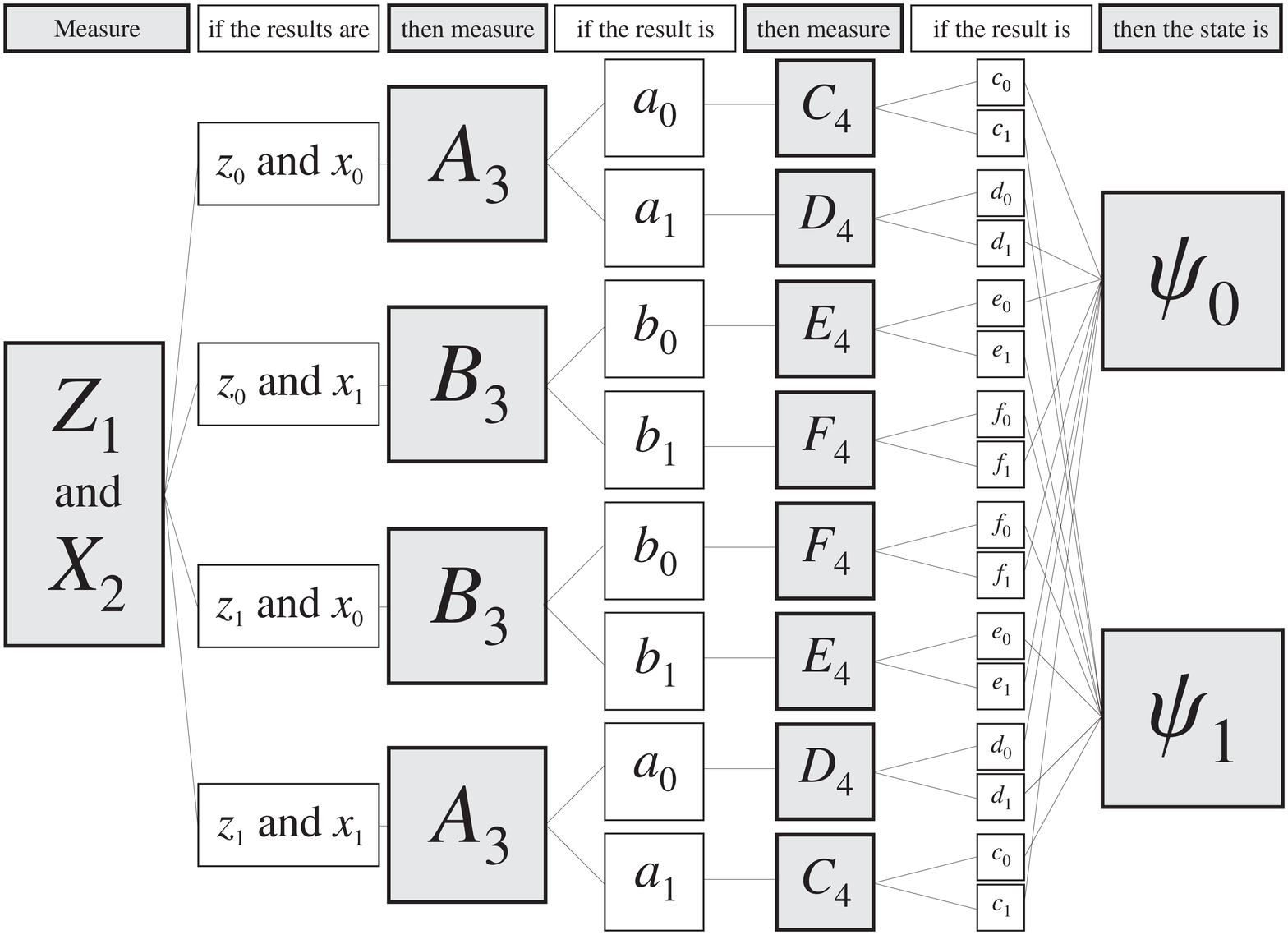}}
\caption{\label{Flow02} Protocol for reliably distinguishing
$|\psi_0\rangle$ and $|\psi_1\rangle$ using a sequence of
single-qubit measurements. Example: first, measure $Z$ on
qubit~$1$ and $X$ on qubit~$2$. If the results are, respectively,
$z_0$ and $x_1$, then measure the observable represented by
$B=|b_0\rangle\langle b_0|-|b_1\rangle\langle b_1|$ on qubit~$3$.
If the result is $b_0$, then measure the observable
$E=|e_0\rangle\langle e_0|-|e_1\rangle\langle e_1|$ on qubit~$4$.
If the result is $e_1$, then the state is $|\psi_1\rangle$.}
\end{figure}

The states $|\psi_0\rangle$ and $|\psi_1\rangle$ are not
distinguishable using {\em fixed} single-qubit measurements.
However, any two orthogonal states are distinguishable by
single-qubit measurements {\em assisted by classical
communication}~\cite{WSHV00}. This means that there is a {\em
sequence} of single-qubit measurements which allows us to reliably
distinguish between $|\psi_0\rangle$ and $|\psi_1\rangle$. In this
sequence, what is measured on one qubit could depend on the result
of a prior measurement on a different qubit. A sequence of
single-qubit measurements which allows us to reliably distinguish
between $|\psi_0\rangle$ and $|\psi_1\rangle$ follows from the
fact that these states can be expressed as
\begin{eqnarray}
|\psi_0\rangle & = & \alpha |z_0,x_0,a_0,c_0\rangle + \beta
|z_0,x_0,a_1,d_1\rangle \nonumber \\ & & + \alpha
|z_0,x_1,b_0,e_0\rangle + \beta |z_0,x_1,b_1,f_1\rangle \nonumber
\\ & & + \beta |z_1,x_0,b_0,f_0\rangle + \alpha
|z_1,x_0,b_1,e_1\rangle \nonumber \\ & & - \beta
|z_1,x_1,a_0,d_0\rangle + \alpha |z_1,x_1,a_1,c_1\rangle,
\label{psi0loc}
\\
|\psi_1\rangle & = & \beta |z_0,x_0,a_0,c_1\rangle + \alpha
|z_0,x_0,a_1,d_0\rangle \nonumber \\ & & + \beta
|z_0,x_1,b_0,e_1\rangle - \alpha |z_0,x_1,b_1,f_0\rangle \nonumber
\\ & & + \alpha |z_1,x_0,b_0,f_1\rangle - \beta
|z_1,x_0,b_1,e_0\rangle \nonumber \\ & & + \alpha
|z_1,x_1,a_0,d_1\rangle - \beta |z_1,x_1,a_1,c_0\rangle,
\label{psi1loc}
\end{eqnarray}
where
\begin{eqnarray}
\alpha & = & {\sqrt{3+\sqrt{6}} \over 2 \sqrt{6}}, \\
\beta & = & {\sqrt{3-\sqrt{6}} \over 2 \sqrt{6}},
\end{eqnarray}
and
\begin{eqnarray}
|a_0\rangle & = & p |z_0\rangle + q |z_1\rangle,\;\;\;\;
|a_1\rangle = q |z_0\rangle - p |z_1\rangle, \\
|b_0\rangle & = & -p |z_0\rangle + q |z_1\rangle,\;\;\;\;
|b_1\rangle = q |z_0\rangle + p |z_1\rangle, \\
|c_0\rangle & = & -r |z_0\rangle + s |z_1\rangle,\;\;\;\;
|c_1\rangle = -s |z_0\rangle - r |z_1\rangle, \\
|d_0\rangle & = & t |z_0\rangle + u |z_1\rangle,\;\;\;\;
|d_1\rangle = u |z_0\rangle - t |z_1\rangle, \\
|e_0\rangle & = & r |z_0\rangle + s |z_1\rangle,\;\;\;\;
|e_1\rangle = s |z_0\rangle - r |z_1\rangle, \\
|f_0\rangle & = & -t |z_0\rangle + u |z_1\rangle,\;\;\;\;
|f_1\rangle = u |z_0\rangle + t |z_1\rangle,
\end{eqnarray}
where
\begin{eqnarray}
p & = & {\sqrt{2-\sqrt{2}} \over 2}, \\
q & = & {\sqrt{2+\sqrt{2}} \over 2}, \\
r & = & {(3+\sqrt{3}) q \over 12 \alpha}, \\
s & = & {(3-\sqrt{3}) q \over 12 \beta}, \\
t & = & {(3-\sqrt{3}) p \over 12 \alpha}, \\
u & = & {(3+\sqrt{3}) p \over 12 \beta}.
\end{eqnarray}
Note that, for instance, the state $|b_0\rangle$ is {\em not}
orthogonal to $|a_0\rangle$ or $|a_1\rangle$. The comparison
between expressions~(\ref{psi0loc}) and~(\ref{psi1loc}) leads us
to a simple protocol for reliably distinguishing between $|\psi_0
\rangle$ and $|\psi_1\rangle$ using a sequence of single-qubit
measurements. This protocol is shown in Fig.~\ref{Flow02}.


\section{Other applications}
\label{Sec5}


\subsection{No-hidden-variables theorems}


Vaidman's aim was to reformulate the GHZ proof of Bell's theorem
into a game ``which can convert laymen into admirers of quantum
theory'' by showing its ``miraculous power''~\cite{Vaidman99}. One
obvious application of the method for always winning Vaidman's
game introduced in this paper is thus to prove Bell's theorem
without inequalities when the local observers do not share any
reference frame. According to Eqs.~(\ref{ZZZ})--(\ref{XXZ}), one
can predict with certainty the value of either ${\cal Z}_j$ or
${\cal X}_j$ (with $j = 1, 2, 3$) from the results of spacelike
separated measurements on the other two four-qubit systems.
Therefore, for any $j$, ${\cal Z}_j$ and ${\cal X}_j$ can be
considered ``elements of reality,'' as defined by Einstein,
Podolsky, and Rosen~\cite{EPR35}. However, it is impossible to
assign predefined values, either $0$ or $1$, to the six
observables ${\cal Z}_j$ and ${\cal X}_j$ satisfying all
predictions given by Eqs.~(\ref{ZZZ})--(\ref{XXZ}).

This proof is of interest, since it shows that a perfect alignment
between the source of entangled states and the local detectors
does not play a fundamental role in Bell's theorem. For instance,
in 1988 Yuval Ne'eman argued that the answer to the puzzle posed
by Bell's theorem was to be found in the implicit assumption that
the detectors were aligned. Ne'eman apparently believed that the
two detectors were connected through the space-time affine
connection of general relativity~\cite{Neeman}. A proof of Bell's
theorem without inequalities and without alignments involving two
observers, eight-qubit states, and only fixed single-qubit
measurements (i.e., without requiring a protocol like the one in
Fig.~\ref{Flow02}) has been introduced in Ref.~\cite{Cabello03c}.
The interest of the proof of Bell's theorem without inequalities
for the state~$|\Psi\rangle$, given in Eq.~(\ref{GHZ12}), and the
local measurements of ${\cal Z}$ and ${\cal X}$, defined
respectively in Eqs.~(\ref{calZ}) and (\ref{calX}), is that such a
proof is valid for 100\% of the events prepared in the
state~$|\Psi\rangle$, instead of only for a small (8\%) subset of
the events in Ref.~\cite{Cabello03c}.

Other interesting application of the state $|\Psi\rangle$ and the
local observables ${\cal Z}$ and ${\cal X}$ is the Kochen-Specker
(KS) theorem of impossibility of noncontextual hidden variables in
quantum mechanics~\cite{KS67}. Mermin showed how the GHZ proof of
Bell's theorem could be converted into a proof of the KS
theorem~\cite{Mermin90d,Mermin93}. Analogously, the proof of
Bell's theorem using $|\Psi\rangle$, ${\cal Z}$, and ${\cal X}$
could be converted into a (subspace-dependent) proof of the KS
theorem, valid even for measurements along imperfectly defined
directions. This is of interest, because it sheds some extra light
on a recent debate about whether or not the KS theorem is still
valid when ideal measurements are replaced by imperfect
measurements~\cite{Meyer99,Kent99,CK00,HKSS99,Mermin99,Appleby02,Cabello02,Larsson02,Breuer02}.


\subsection{Reducing the communication complexity with prior entanglement}


Vaidman's game can also be seen as a scenario in which the
communication complexity of a certain task can be reduced if the
players are allowed to share some prior entangled state. In
Vaidman's game the task is to always win the game. Without quantum
resources, this task requires at least one of the players to send
$1$~bit to other player after the question ($Z$ or $X$) has been
posed to him. However, if they initially share a GHZ state, the
task does not require any transmission of classical information
between the players.

A similar example of reduction of the communication complexity
needed for a task if the parties share a GHZ state was discovered
by Cleve and Buhrman~\cite{CB97}, reformulated by Buhrman {\em et
al.}~\cite{BvHT99}, and attractively presented by Steane and van
Dam~\cite{Sv00} as follows: a secret integer number $n_A+n_B+n_C$
of apples, where $n_j=0$, $\frac{1}{2}$, $1$, or $\frac{3}{2}$, is
distributed among three players, Alice, Bob, and Charlie, of the
same team. Each of them is in an isolated booth. The team wins if
one of the players, Alice, can ascertain whether the total number
of distributed apples is even or odd. The only communication
allowed is that each of the other two players can send $1$~bit to
Alice after seeing the number of apples each of them got. Assuming
that each of the~$32$ possible variations of apples occurs with
the same probability and using only classical communication, Alice
cannot guess the correct answer in more than~75\% of the cases.
However, the players can always win if each has a qubit of a trio
prepared in the state $|{\rm GHZ}\rangle$ given in
Eq.~(\ref{GHZ03}), and each player~$j$ applies to his qubit the
rotation
\begin{equation}
R(n_j)=|y_0\rangle\langle y_0|+e^{i n_j \pi}|y_1\rangle\langle
y_1|,
\end{equation}
where $n_j$ is his number of apples, and then measures the spin of
his qubit along the $z$~direction. Finally, Bob and Charlie send
their outcomes to Alice. The success of the method is guaranteed
by the following property:
\begin{eqnarray}
\lefteqn{R(n_A) \otimes R(n_B) \otimes R(n_C) |{\rm GHZ}\rangle =} \nonumber \\
& & \left\{\begin{array}{ll}
|{\rm GHZ}\rangle & \mbox{if $n_A+n_B+n_C$ is even} \\
|{\rm GHZ}^\perp\rangle & \mbox{if $n_A+n_B+n_C$ is odd,}
\end{array} \right.
\end{eqnarray}
where
\begin{eqnarray}
|{\rm GHZ}^\perp\rangle & = & {i \over 2} (|z_0,z_0,z_1\rangle+|z_0,z_1,z_0\rangle \nonumber \\
& & +|z_1,z_0,z_0\rangle-|z_1,z_1,z_1\rangle),
\label{GHZ03perp}
\end{eqnarray}
can be reliably distinguished from $|{\rm GHZ}\rangle$ by local
measurements along the $z$~direction. This method assumes that all
players {\em share a reference frame} during the protocol.
However, such an assumption is not needed if each player replaces
his qubit belonging to a trio prepared in~$|{\rm GHZ}\rangle$ by
four qubits belonging to a dozen prepared in~$|\Psi\rangle$. The
local operations [i.e., the rotation $R(n_j)$ and the measurement
along the $z$~direction] are replaced by a protocol, using only
single-qubit measurements, for reliably distinguishing between two
four-particle states which are invariant under $U_j \otimes U_j
\otimes U_j \otimes U_j$.


\subsection{Quantum cryptography}


Other application in which the use of GHZ states provides
advantages over any classical protocol is the secret sharing
scenario~\cite{HBB99,KKI99,TZG99,Cabello00}: Alice wishes to
convey a cryptographic key to Bob and Charlie in such a way that
they both can read it only if they cooperate. In addition, they
wish to prevent any eavesdropper from acquiring any information
without being detected. It is assumed that the players share no
previous secret information nor any secure classical channel but,
although it is not usually explicitly stated, it is assumed that
all three parties {\em share a reference frame}. Once more, such a
requirement can be removed if we replace the GHZ state with the
state~$|\Psi\rangle$, and the measurements of~$Z$ and~$X$ with
measurements of ${\cal Z}$ and ${\cal X}$.


\subsection{Conclusion}


To sum up, the interest in rotationally invariant states (i.e.,
those invariant under $U \otimes \ldots \otimes U$, where $U$ is a
unitary operation) goes beyond their use for decoherence-free
fault-tolerant universal quantum
computation~\cite{KBLW01,ZR97a,LCW98,BEGKCW03}, solving the
Byzantine agreement problem~\cite{FGM01,Cabello02b,Cabello03b},
and transmitting classical and quantum information between parties
who do not share a reference frame~\cite{BEGKCW03,BRS03}. {\em
Entangled} rotationally invariant states (i.e., those invariant
under $U_A \otimes \ldots \otimes U_A \otimes \ldots \otimes U_N
\otimes \ldots \otimes U_N$), like the state $|\Psi\rangle$ given
in Eq.~(\ref{GHZ12}), can be used to overcome certain assumptions in
the proofs of nonexistence of hidden variables, can be applied to
reduce the communication complexity of certain tasks, even if the
parties do not share any reference frame, and to distribute secret
keys among parties who do not share unspeakable information.


\begin{acknowledgments}
This paper was sparked by a question raised by H.~Weinfurter about
the possibility of developing a GHZ-like proof of Bell's theorem
involving observers who do not share a reference frame and using
only single-qubit measurements. I would like to thank him and
M.~Bourennane for useful discussions on this and related subjects,
N.~D.~Mermin and A.~Peres for their comments on a preliminary
version, and I.~Garc\'{\i}a-Pelayo for sending me the proofs of
Ref.~\cite{GG03}. This work was supported by the Spanish
Ministerio de Ciencia y Tecnolog\'{\i}a Project No.~BFM2002-02815
and the Junta de Andaluc\'{\i}a Project No.~FQM-239.
\end{acknowledgments}



\end{document}